\documentclass[journal=ancac3,manuscript=article,etalmode=truncate,maxauthors=100]{achemso}

\usepackage[version=3]{mhchem} % Formula subscripts using \ce{}
\usepackage{fixltx2e}
\usepackage{textcomp}
\usepackage{xcolor}
\usepackage{soul}
\usepackage{dirtytalk}
\usepackage{microtype}
\usepackage{verbatim}

\setkeys{acs}{etalmode=truncate,maxauthors=10}

\author{Alvaro Rodriguez}
\affiliation[JHI]
{J. Heyrovský Institute of Physical Chemistry, Czech Academy of Sciences, Dolejškova 2155/3, 182 23 Prague, Czech Republic}

\author{Martin Kalbac}
\affiliation[JHI]
{J. Heyrovský Institute of Physical Chemistry, Czech Academy of Sciences, Dolejškova 2155/3, 182 23 Prague, Czech Republic}

\author{Otakar Frank}
\affiliation[JHI]
{J. Heyrovský Institute of Physical Chemistry, Czech Academy of Sciences, Dolejškova 2155/3, 182 23 Prague, Czech Republic}
\email{otakar.frank@jh-inst.cas.cz}

\title[Strong Localization Effects in the Photoluminescence of Transition Metal Dichalcogenide Heterobilayers]
  {
  Strong Localization Effects in the Photoluminescence of Transition Metal Dichalcogenide Heterobilayers}
  

\begin{document}

%\begin{tocentry}

%\includegraphics{figures/toc}

%\end{tocentry}

\begin{abstract}
The emergence of various exciton-related effects in transition metal dichalcogenides (TMDC) and their heterostructures has inspired a significant number of studies and brought forth several possible applications. Often, standard photoluminescence (PL) with microscale lateral resolution is utilized to identify and characterize these excitonic phenomena, including interlayer excitons (IEXs). We studied the local PL signatures of van der Waals heterobilayers composed of exfoliated monolayers of the (Mo,W)(S,Se)$_2$ TMDC family with high spatial resolution (down to 30~nm) using tip-enhanced photoluminescence (TEPL) with different orders (top/bottom) and on different substrates. We evidence that other PL signals may appear near the reported energy of the IEX transitions, possibly interfering in the interpretation of the results. While we can distinguish and confirm the presence of IEX-related PL in MoS$_2$-WS$_2$ and MoSe$_2$-WSe$_2$, we find no such feature in the MoS$_2$-WSe$_2$ heterobilayer in the spectral region of 1.7--1.4~eV, where the IEXs of this heterobilayer is often reported. We assign the extra signals to the PL of the individual monolayers, in which the exciton energy is altered by the local strains caused by the formation of blisters and nanobubbles, and the PL is extremely enhanced due to the decoupling of the layers. We prove that even a single nanobubble as small as 60~nm---hence not optically visible---can induce such a suspicious PL feature in the micro-PL spectrum of an otherwise flat heterobilayer. 
\end{abstract}

%\section{Introduction}
\newpage

The fabrication of two-dimensional (2D) materials by stacking different layered materials on top of each other has attracted the attention of the scientific community due to the unique properties and the potential implementation options of the thus created van der Waals heterostructures in several optoelectronic devices, such as photodetectors, light-emitting diodes, or lasers.\cite{Geim2013,Novoselov2016,Jariwala2016,Long2019,Ricciardulli2020,Liueaav2019,Chhowalla2016} The understanding of the fundamental aspects that govern the properties of these systems is essential for the development of the targeted applications. Among 2D materials, transition metal (Mo, W) dichalcogenides (S, Se) (TMDCs) have emerged as research topics because of their semiconductor nature with direct bandgap energy in the visible region at the monolayer limit.\cite{Mak2010,Splendiani2010,Tongay2013} Heterobilayers fabricated by the stacking of different monolayers exhibit strong interlayer coupling in a type-II band alignment and potentially a radiative recombination of interlayer excitons (IEXs) when the photoexcited electron and hole are localized in separate layers.\cite{Rivera2018} IEXs formed in TMDC heterobilayers possess unique properties, such as high binding energies, long radiative lifetimes, electrical tunability, and valley polarization, which allows the excitation of either electrons or holes in a particular valley using circularly polarized light.\cite{Rivera2016,Ciarrochi2019,Wei2018,Jin2019,Seyler2019}

Experimentally, IEX emissions have been detected at room temperature by using $\mu$-photoluminescence ($\mu$-PL) in several combinations of TMDCs. For example, the IEX emissions have been observed in MoS$_2$-WS$_2$ and MoS$_2$-WSe$_2$ in the energy range between 1.4--1.6~eV and 1.3--1.4~eV, respectively.\cite{Gong2014,Tongayb2014,Okada2018,Kiemle2020,Hanbicki2018,Nayak2017,Rivera2018b,Nagler2017} The MoS$_2$-WSe$_2$ heterobilayer presents a special case due to the large lattice mismatch and the offset of energy levels.\cite{Ozelik2016} The IEX in MoS$_2$-WSe$_2$ was reported to appear at energies between 1.6--1.55~eV, and it was assigned to the momentum-space indirect interlayer emission.\cite{Kunstmann2018,Nagler2019} For this system, Fang \textit{et al.}\cite{Fang2014} first reported the IEX at $\approx$~1.55~eV, while Kunstmann \textit{et al.}\cite{Kunstmann2018} reported similar PL features at 1.6~eV. Although SiO$_2$ substrates were used in both studies, the samples differed in the dry transfer technique. In the former work,\cite{Fang2014} the exfoliated monolayers were picked up from the substrate with polydimethylsiloxane (PDMS), whereas the monolayers were exfoliated directly on PDMS in the latter work.\cite{Kunstmann2018} Several researchers have reported similar PL features in samples prepared by a variety of methods, for example, using PDMS stamps for transferring only the top layer, or after encapsulating the layer with hBN, or even when using a wet transfer method, all in the energy range of 1.6--1.5~eV.\cite{Chiu2014,Unuchek2018,Ji2020,Lin2015} However, a few recent authors have argued the presence of the MoS$_2$-WSe$_2$ space indirect interlayer emission further in the infrared region, at $\approx$1.1~eV.\cite{Huang2020,Karni2019}

Despite reported observations of interlayer emissions in various systems by conventional $\mu$--PL methods, studies of local PL signatures and the effects of heterogeneities on the interaction between the layers are lacking. This is particularly important when preparing heterobilayers by other means than direct growth. In manually stacked heterostructures, the formation of various out-of-plane features is unavoidable, in part due to contamination and in part due to deformation induced by the handling. Often, the resulting topography imperfections cannot be spotted in the diffraction limit, be it optical microscopy or spectroscopy. \cite{Pizzocchero2016,Schwartz2019} In previous works, tip-enhanced photoluminescence (TEPL) has been utilized to probe local nanoscale heterogeneities in different TMDC monolayers.\cite{Park2016,Rahaman2017,Okuno2018,Rodriguez2019,Tim2020} Herein, we catalogue the local PL signatures of variously stacked TMDC heterobilayers and confirm the presence of interlayer emission from the flat parts of MoS$_2$-WS$_2$ and MoS$_2$-MoSe$_2$. In contrast, no interlayer emission in the range of 1.5--1.6~eV takes place in WSe$_2$-MoS$_2$ heterobilayers with stacking angles close to 0º, even in cases where a perfect interface is documented by a clear Moiré pattern in topography. We identify different surface imperfections by TEPL spectroscopy in  WSe$_2$-MoS$_2$ heterobilayers with emission energies ranging from 1.6~eV to 1.5~eV and assign those bands to the PL of the top layer, which shifts due to the strain caused by the encapsulated contamination.

\section{Results}

\begin{figure}
 \includegraphics%[width\=textwidth]
 {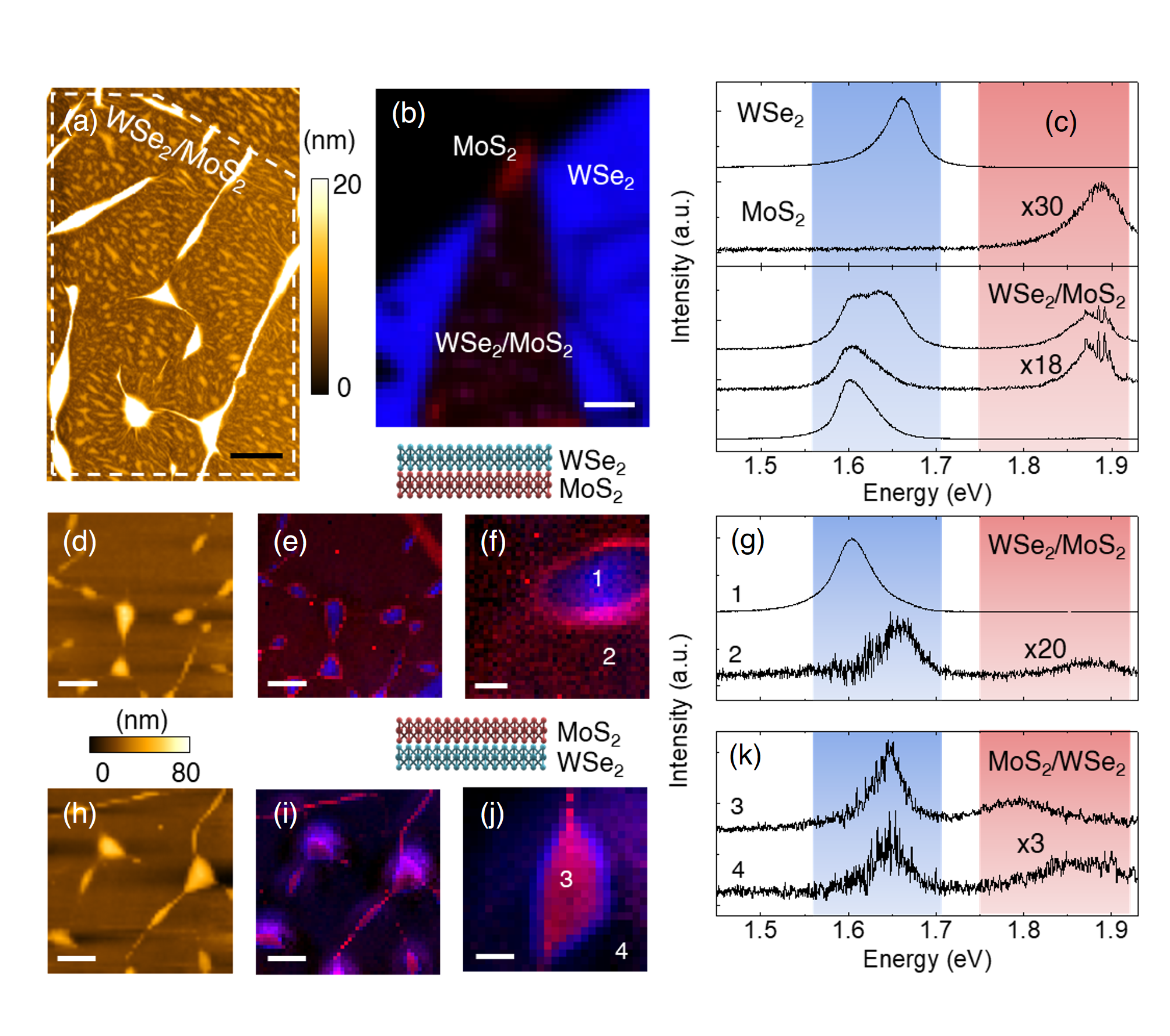}
  \caption{\textbf{Comparison between micro and nano PL of WSe$_2$-MoS$_2$ heterobilayers on SiO$_2$.} (a) Representative topography image of a WSe$_2$/MoS$_2$ heterobilayer. (b) A $\mu$-PL intensity map of WSe$_2$/MoS$_2$ showing strong PL quenching in the area where the monolayers are interacting. Red and blue colors refer to the integrated intensity in the energy range indicated in (c). (c) Normalized $\mu$-PL spectra of WSe$_2$, MoS$_2$, and WSe$_2$/MoS$_2$. (d--k) Nano optical characterization of WSe$_2$-MoS$_2$ heterobilayers: (d) topography, (e--f) TEPL intensity maps, and (g) normalized TEPL spectra of WSe$_2$/MoS$_2$ taken at selected points indicated in (f). (h) Topography, (i--j) TEPL intensity maps and (k) normalized TEPL spectra of MoS$_2$/ WSe$_2$ at selected points indicated in (j). Scale bars: 1~$\mu$m in (a), 5~$\mu$m in (b), 1~$\mu$m in (d-e) and (h-i), and 200~nm in (f) and (j). Pixel sizes: 100~nm in (e) and (i), 30~nm in (f), and 40~nm in (j).}
  \label{fgr:1}
\end{figure}

We prepared heterobilayers by combining different mechanically exfoliated TMDC monolayers by a dry transfer method (see Experimental for details). Figure \ref{fgr:1}(a) shows the topography of a WSe$_2$/MoS$_2$ heterobilayer on SiO$_2$/Si, illustrating the typical heterogeneities introduced by the stacking of 2D materials. Note that throughout the text, the heterobilayers are always designated as A/B, where A is the top layer and B the bottom one. If the heterobilayer is labeled as A-B, we do not make a distinction of the layer order. Besides small bubbles created in the bottom layer (MoS$_2$), the topography is dominated by the formation of blisters in the top layer (WSe$_2$) due to the lateral diffusion of contamination, leading to a clean interface in between the blisters.\cite{Kretinin2014} Otherwise, the top layer maintains the topography of the bottom layer. The PL in the heterobilayer area is strongly quenched, owing to the strong interlayer charge transfer between the two monolayers,\cite{Chiu2015} as shown in Figure \ref{fgr:1}(b). The varied $\mu$-PL signals within the heterobilayer area are outlined in the bottom panel of Figure \ref{fgr:1}(c), whereas the PL spectra in the top panel correspond to the individual monolayers. The heterobilayer exhibits emissions at 1.65 and 1.85~eV, corresponding to the intralayer excitons for WSe$_2$ and MoS$_2$, respectively. An extra band is centered at $\approx$1.6~eV, and it becomes weaker in the regions where the atomic force microscopy (AFM) topography image looks more flat. This band is within the range for the reported momentum-space indirect interlayer emission in WSe$_2$/MoS$_2$,\cite{Kunstmann2018} and it might be assigned to such excitonic species. However, the limited resolution of the $\mu$-PL system makes it impossible to discern between signals coming from the nano-sized features observed in the AFM image. To improve the spatial resolution of the PL map, we performed a TEPL survey in the same sample area, and the results are summarized in Figure \ref{fgr:1}(d--g). When comparing the AFM image and the TEPL map measured during the same scan, one can see that the PL signal from the blisters is stronger than the PL observed in the flat areas. Moreover, the spectra extracted from the enlarged map reveal that the PL of the blisters is redshifted by 0.05~eV with respect to the WSe$_2$ neutral intralayer emission (Figure \ref{fgr:1}(g) and S6). This band coincides with the position of the extra band observed by $\mu$--PL, and it is often assigned in the literature to the IEX.\cite{Kunstmann2018,Fang2014,Unuchek2018,Chiu2014} We note that we did not take into consideration the twisted angle since the emission at 1.6~eV was observed for all the angles studied in the previous works.\cite{Kunstmann2018,Nagler2019}

Since the TEPL from the flat areas in Figure \ref{fgr:1}(g) did not show any additional band other than the emission from the individual monolayers, we prepared samples with an inverse layer order by placing the MoS$_2$ on top (see Figure S3 for $\mu$--PL maps). In MoS$_2$/WSe$_2$, the TEPL map (Figure \ref{fgr:1}(h--k)) shows behavior opposite to the WSe$_2$/MoS$_2$ case, and now the PL signal from the blisters shifts by 0.05~eV with respect to the MoS$_2$ intralayer emission (from 1.85~eV to 1.8~eV), while no extra band is observed at $\sim$1.6~eV. Hence, in both cases, the blisters form in the top layers, and their intrinsic PL signal redshifts while the flat areas do not show any extra bands. The more intense PL of the layers in the blisters suggests a weaker interlayer interaction takes place in those areas. We attribute the observed redshift to the strain caused by the formation of contamination pockets.\cite{Aslan2018} The exact position of the emission will depend on the level of strain, in a fashion similar to the case of monolayer WS$_2$ on a gold substrate, as reported recently.\cite{Darlington_2_2020} We point out that the PL from the small bubbles formed between the bottom TMDC monolayer and the SiO$_2$ substrate (observed in the higher resolution AFM image in Figure \ref{fgr:1}(a)) seems to be unaffected. 

\begin{figure}
 \includegraphics%[width=0.5\textwidth]
 {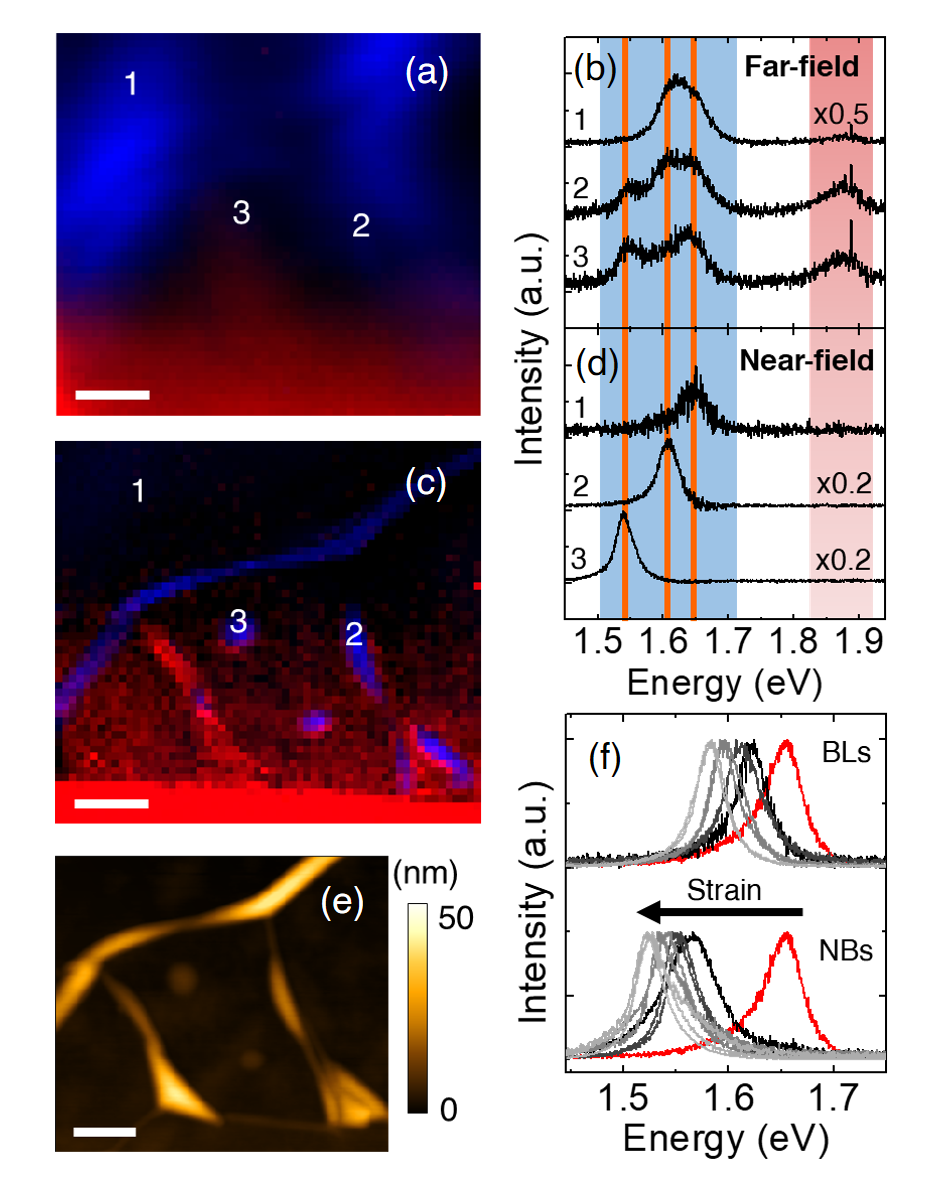}
  \caption{\textbf{Comparison between micro and nano PL of WSe$_2$/MoS$_2$ on hBN.} (a) Far-field intensity map and (b) selected far-field spectra. (c) Near-field intensity map and (d) selected near-field spectra. Pixel size is 50~nm. (e) AFM topography of the same area. (f) PL spectra of heterogeneities separated in two groups: blisters (BLs) and nanobubbles (NBs). Scale bar is 500~nm.}
  \label{fgr:2}
\end{figure}

To shed more light on the possible disentanglement of local strain-induced shifts and interlayer emission (if it exists) in this particular heterobilayer, we prepared the same structure but on hexagonal boron nitride (hBN) to assess the effect of the substrate. For these samples, we expect that blisters will appear in both MoS$_2$ and WSe$_2$ layers, independently of the stacking order. Figure \ref{fgr:2}(a) shows the far-field PL intensity map of the heterobilayer on hBN when WSe$_2$ is placed on top of the MoS$_2$. The far-field here is taken with side illumination in our TEPL setup \cite{Rodriguez2019}. As expected, a more heterogeneous PL distribution is observed due to the formation of both MoS$_2$ and WSe$_2$ blisters. The spectra in Figure \ref{fgr:2}(b) shows the far-field PL spectra of selected areas in the heterostructure, where different bands near the WSe$_2$ emission of the neutral exciton can be observed. Interestingly, the spectral features acquired in the far-field can be resolved in the near-field map (Figure \ref{fgr:2}(c)). Figure \ref{fgr:2}(d) shows three different bands, making the assignment straightforward. Number one is for the intrinsic WSe$_2$ intralayer emission at 1.65~eV, number two is similar to the previously observed band for the blisters at 1.6~eV, and number three at 1.55~eV exhibits an even larger shift. The topography image of the same sample area (Figure \ref{fgr:2}(e)) shows that the intensity of the TEPL signals correlates with the topography features observed in the AFM image. Interestingly, besides the previously described blisters (\#2 in (c) and (d)), we observed another feature with a more circular shape (\#3) which corresponds to the PL emission at 1.55~eV. Such features have recently been observed in WSe$_2$/hBN heterostructures, introduced as nanobubbles, \cite{Darlington_1_2020} and assigned to the emission from localized exciton (LX) states. We analyzed the emission energies of the various surface heterogeneities for different samples and separated them into two groups: blisters and nanobubbles (Figure \ref{fgr:2}(f)). In general, nanobubbles (1.52--1.58~eV) induce larger PL shift than blisters (1.58--1.62~eV) due to larger strain and localization. The shapes of both types indicate they are filled with gas or liquid, rather than with a nanoparticle (cf. Figure S11).\cite{Khestanova2019} 

\begin{figure}
 \includegraphics[width=0.5\textwidth]
 {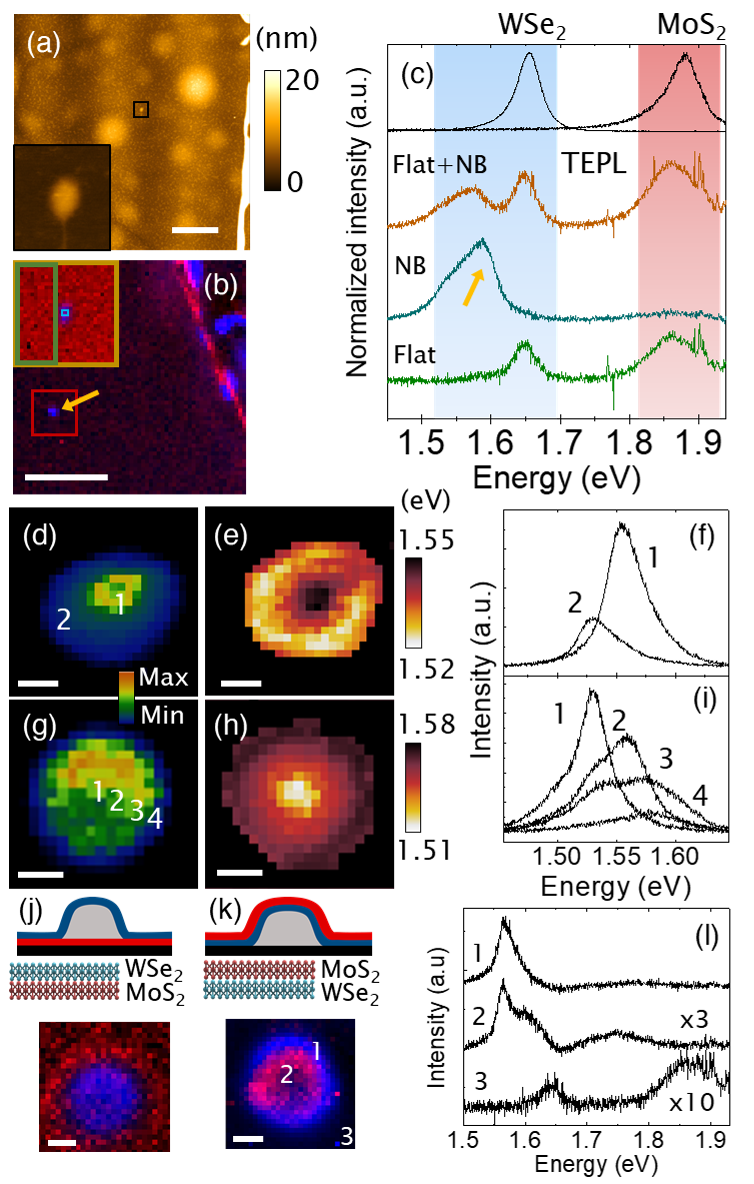}
  \caption{\textbf{Characterization of nanobubbles in blister-free WSe$_2$/MoS$_2$ on hBN.} (a) Topography image and (b) TEPL intensity map. (c) Comparison between TEPL spectra obtained as the average of all the spectra acquired in the areas indicated in (b) with matching colors. TEPL intensity and energy maps of small (d--e) and large (g--h) WSe$_2$ nanobubbles on MoS$_2$. (f) and (i) TEPL spectra from the small and large nanobubbles, respectively, taken at the points indicated in (d) and (g). The big nanobubble shows multiple local states. TEPL intensity maps of nanobubbles in (j) WSe$_2$/MoS$_2$ and (k) MoS$_2$/WSe$_2$. (l) TEPL spectra of the MoS$_2$/WSe$_2$ nanobubble at selected positions as indicated in (k). Scale bars: 1~$\mu$m (a, b), 50~nm (d, e), and 100~nm (g, h, j, k). Pixel sizes: 50~nm (b), 20~nm (g, h, j), 15~nm (k), and 10~nm (d, e).}
  \label{fgr:3}
\end{figure}

Besides the emissions from the blisters and nanobubbles, none (\textit{e.g.}, from interlayer excitons) were observed in the spectra. To evaluate the effect of \say{optically invisible} nanobubbles on the resulting PL spectra, we measured a TEPL map in an area that contains no visible blisters and is large enough to be resolved with conventional $\mu$--PL methods. Figure \ref{fgr:3}(a) shows a 4~$\times$~4~$\mu$m$^2$ \say{flat} area of the same type of heterostructure as in Figure \ref{fgr:2} (\textit{i.e.}, WSe$_2$/MoS$_2$ on hBN). Apart from the bottom bubbles originated between the hBN layer and the SiO$_2$ substrate, blisters are not created in that area, and the two TMDC layers are copies of the topography of the hBN, which suggests that a clean interface between MoS$_2$ and WSe$_2$ should be generated. The TEPL map of that area (Figure \ref{fgr:3}(b)) shows a uniform signal distribution across the area without blisters with the exception of a strong PL feature (marked with a yellow arrow) arising from a small nanobubble in the center of the area. This particular nanobubble can be identified in the AFM image (Figure \ref{fgr:3}(a). The nanobubble is notably smaller than the bottom hBN bubbles, which, however, do not provide any recognizable PL change. 

 The average TEPL spectra of the areas indicated in the inset of Figure \ref{fgr:3}(b) are displayed in Figure \ref{fgr:3}(c) in the respective colors. Three bands are observed in the brown spectrum, which corresponds to the average of all the spectra in the inset. Besides the intralayer emissions at 1.65~eV and 1.85~eV for WSe$_2$ and MoS$_2$, respectively, an extra band is observed at 1.58~eV, which could immediately be assigned to the interlayer emission, according to the reported IEX energy in some works. Nevertheless, the high spatial resolution achieved with TEPL allowed us to resolve those bands, revealing that the emission at 1.58~eV is actually the emission from the nanobubble, while the flat areas exhibit only the MoS$_2$ and WSe$_2$ intralayer excitons.

Figure \ref{fgr:3}(d--i) shows the PL of different WSe$_2$ nanobubbles (on MoS$_2$) measured with TEPL. Since the PL is extremely enhanced in those areas, it can be nicely resolved and enables the observation of the particular doughnut-like distribution of PL as a consequence of the higher strain in the nanobubbles periphery (Figure (\ref{fgr:3}(e)).\cite{Carmesin2019,Darlington_1_2020} In the spectra shown in Figure \ref{fgr:3}(f), one can see the more intense, less shifted PL band coming from the center of the nanobubble, while the less intense PL band with the larger shift can be observed along the circumference. In contrast, the big nanobubble (Figure \ref{fgr:3}(h)) shows a different PL energy distribution with multiple local strain states across the nanobubble, as observed in the PL spectra (Figure \ref{fgr:3}(i)). The localized states are spread into concentric rings in the nanobubble. The periphery is now dominated by higher energy states due to a weaker confinement potential in those regions. The non-trivial shape of spectrum 1 in Figure \ref{fgr:3}(i) suggests that an even better lateral resolution would be needed to fully resolve the varying localized strain states. The multitude of relatively sharp emissions from the nanobubble supports the idea of a strong relation between these heterogeneities and quantum emitters observed in WSe$_2$,\cite{Tonndorf2015,Shepard_2017,Branny2017,Palacios2017} as suggested by Darlington \textit{et al.}\cite{Darlington_1_2020} 

An interesting effect occurs when placing an extra layer on top of a WSe$_2$ nanobubble or blister, as in MoS$_2$/WSe$_2$ heterobilayers. Figure \ref{fgr:3}(j--k) shows the TEPL intensity map of nanobubbles for the two main configurations encountered in this work. As previously shown (Figure \ref{fgr:2}), when the WSe$_2$ monolayer is on top, the PL from the WSe$_2$ nanobubble is redshifted (Figures \ref{fgr:2}(d) and \ref{fgr:2}(f)) and the PL signal of MoS$_2$ does not shift, but it is more intense in the areas immediately surrounding the nanobubble. When a MoS$_2$ layer is placed on top of WSe$_2$, the distribution of the PL intensity changes, exhibiting a higher PL intensity at the periphery. The \say{bottom} WSe$_2$ nanobubble (\textit{i.e.}, below the MoS$_2$ layer) is disturbed by the top layer modifying the strain distribution. In the center of the nanobubble, both layers exhibit strain-induced PL shifts (spectrum 2 in Figure \ref{fgr:3}(l)); however, the PL band of WSe$_2$ is now less shifted, but weaker and broader, similarly to the effects observed in the big nanobubbles. We highlight the importance of the layer order-dependent nanobubble strain since the thus induced difference in PL shifts can be entangled with possible charge-doping effects if IEX emission is present.\cite{Ji2020}    

We also noticed a considerable difference between nanobubbles formed in MoS$_2$ and WSe$_2$ monolayers. When comparing nanobubbles of the same size and on the same substrate (\textit{e.g.}, on hBN), thereby assuming similar strain levels reached in the nanobubble, the TEPL signal of MoS$_2$ is weak and the PL shift is similar to that observed in the MoS$_2$ blisters with a much wider spectral line shape (Figure S8). Moreover, we measured larger PL shifts for nanobubbles in WSe$_2$ (up to 150~meV with respect to the neutral excitons) than in MoS$_2$. This is consistent with the assignment of the emission from the nanobubble to LX states, since localized states in MoS$_2$ are observed as a broad PL band, while localized states in WSe$_2$ exhibit a series of sharp bands. \cite{Wang2014,Jadczak_2017,Linhart2019}

Concerning the blisters, where the shift seems to be universal for all studied heterobilayers, it is plausible that the shift is also a consequence of strain. The assignment of the bands to different excitonic species can be more ambiguous. For WSe$_2$, dark exciton emission was observed at 1.6~eV,\cite{Park2018} matching well with the energies at which we observed the emission of the blisters. However, the dark excitons were observed when using Au as a substrate, forming a nano-cavity responsible for the enhanced emission. In our samples, it appears unlikely that dark excitons can be observed in the strained areas at room temperature.\cite{Zhang2015} We also exclude the trion as being responsible for the measured low-energy band because the shift of the band is notably larger than the reported energy difference between the trion and neutral exciton emissions.\cite{Li2018} 

\begin{figure}
 \includegraphics%[width\=textwidth]
 {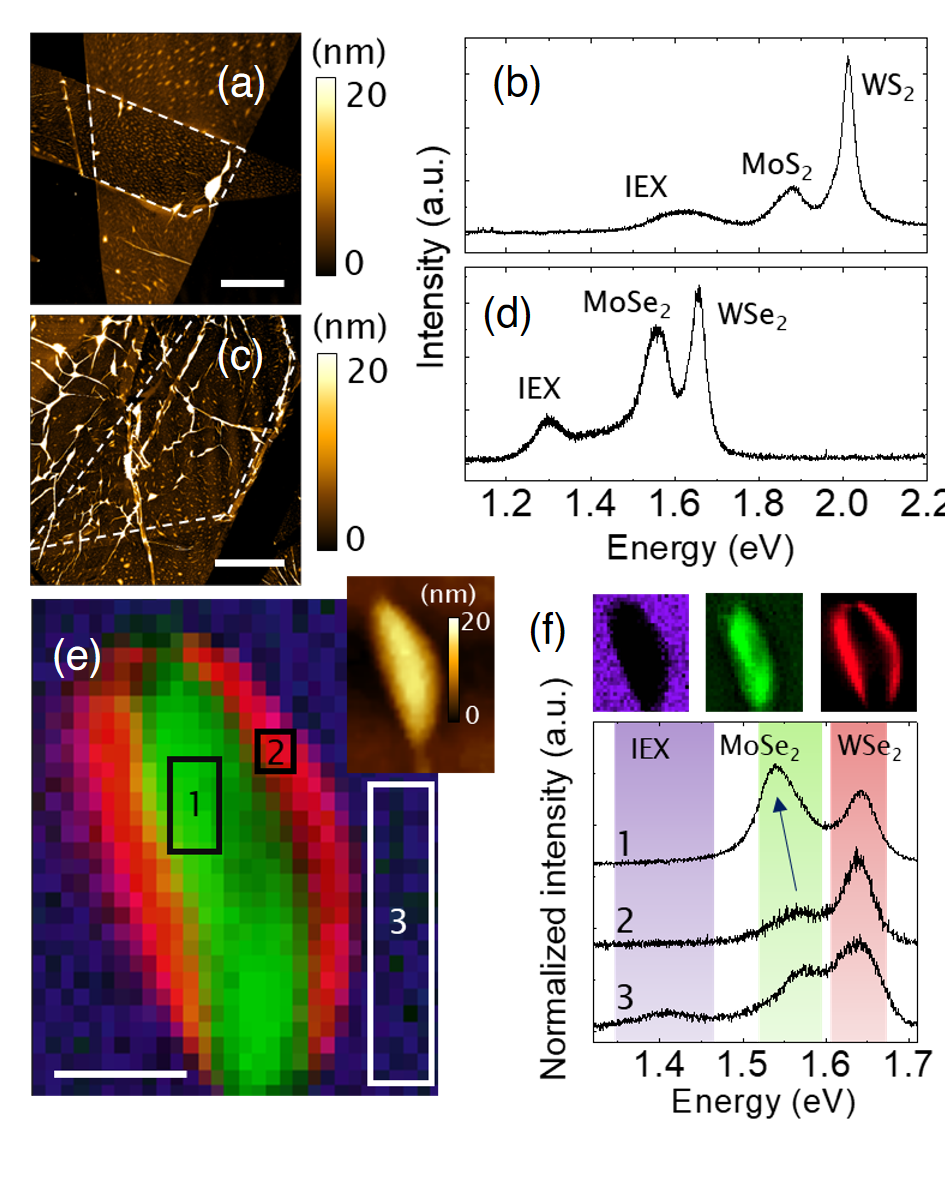}
  \caption{\textbf{Interlayer emission from different heterobilayers on SiO$_2$.} (a) Topography and (b) $\mu$-PL spectrum of MoS$_2$/WS$_2$. (c) Topography and (d) $\mu$-PL spectrum of MoSe$_2$/WSe$_2$. (e) TEPL intensity map of a MoSe$_2$/WSe$_2$ blister, 25~nm pixel size, 1s acquisition time. Colors in the map correspond to the integrated intensity in the energy ranges indicated in (f). The inset shows the corresponding AFM image. (f) TEPL spectra of the areas indicated in (e). Top panel shows the individual intensity map in each energy range. Interlayer emission takes place only in the flat areas. Scale bars: 5~$\mu$m (a, c) and~200 nm (e).}
  \label{fgr:4}
\end{figure}

To assess if the absence of measurable IEX emission in MoS$_2$-WSe$_2$ heterobilayers (regardless of order) is due to our sample preparation method, we prepared heterobilayers of different systems, such as MoS$_2$-WS$_2$ and MoSe$_2$-WSe$_2$ on SiO$_2$/Si substrates (Figure \ref{fgr:4}) using the same assembly method. Figure \ref{fgr:4} (a--d) shows the $\mu$-PL spectra obtained in both samples when placing the MoX$_2$ layer on top. Both intralayer and interlayer excitons can be observed, consistent with previous reports.\cite{Tongayb2014,Gong2014,Okada2018,Kiemle2020,Hanbicki2018, Nagler2017, Nayak2017, Rivera2018b}

We also observed interlayer emission when placing the WX$_2$ layer on top (Figures S4 and S5). Again, the large PL shifts caused by the blisters and nanobubbles produce characteristic shapes in the spectra. Figure \ref{fgr:4}(e--f) shows the TEPL map of a blister created in a MoSe$_2$/WSe$_2$ heterobilayer. The color map in Figure \ref{fgr:4}(e) shows the integrated intensity for different energy ranges indicated in Figure \ref{fgr:4}(f). Similarly to the MoS$_2$/WSe$_2$ heterobilayer, the PL from the blister exhibits a strain-dependent redshift of the neutral exciton from the top layer (green range), while the PL of the bottom layer is more intense in the blister surroundings (red). The PL band of the IEX can only be observed in the flat areas (magenta) due to the strong interaction between the layers.

\begin{figure}
 \includegraphics[width=\textwidth]
 {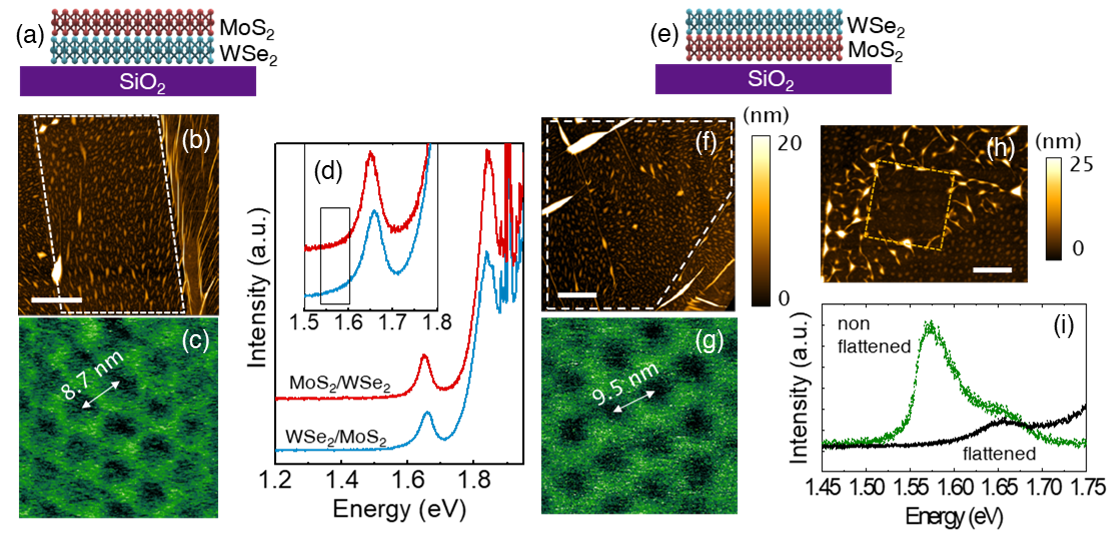}
  \caption{\textbf{Characterization of blister-free MoS$_2$/WSe$_2$ and WSe$_2$/MoS$_2$ heterobilayers on SiO$_2$.} (a, e) Sketch of the sample (b, f) topography images showing mostly the bottom bubbles. The twisted angle close to 0º allows us to measure the Moiré patterns with a standard AFM scan (c, g) showing a periodicity of $\sim$~9 nm. (d) The $\mu$--PL spectra of the samples showing absence of emission in the range of 1.55--1.6 eV. (h) Topography image of a WSe$_2$/MoS$_2$ heterobilayer on SiO$_2$ after flattening the square area (indicated in the figure with a yellow line) with an AFM tip. (i) PL spectra of the area before (green line) and after the flattening (black line). Scale bar is 2~$\mu$m.}
  \label{fgr:5}
\end{figure}

Once we had excluded our sample preparation method as the reason for not detecting the IEX in MoS$_2$-WSe$_2$, we focused on preparing these heterobilayers with the strongest possible interlayer coupling. Figures \ref{fgr:5}(a--g) show a large clean area (4~$\times$~4~$\mu$m$^2$) of two heterobilayers with MoS$_2$ (a--c) and WSe$_2$ (e--g) on top. The AFM images in Figures \ref{fgr:5}(b, f) show the presence of numerous bubbles; however, those are created when placing the first layer on the substrate, and they do not have any effect on the PL. The second layer copies the topography of the first one. We were able to image a Moiré pattern of small twist angles in both samples (Figures \ref{fgr:5}(c, g)) with simple scans in Peak Force Tapping mode, confirming a good interaction and clean interface between the layers.\cite{Zhang2017,Waters2020} The $\mu$-PL spectra displayed in Figure \ref{fgr:5}(d) shows two almost identical curves, absent of any extra band in the reported IEX emission range. On top of that, the bands corresponding to the individual intralayer excitons are of the same energy in spite of the reversed layer order.   

Finally, we tried to see the IEX emission in the heterobilayers both on SiO$_2$ and hBN after removing the blisters by applying the nano-squeegee technique (scanning the area with an AFM tip in contact mode with a setpoint large enough to considerably press the sample, but small enough to avoid damaging it).\cite{Rosenberger2018} Figures \ref{fgr:5}(h--i) show the flattened area and the corresponding PL spectra taken before and after cleaning a WSe$_2$/MoS$_2$ heterobilayer on SiO$_2$. The emission at 1.6~eV completely vanishes after the removal of the blisters, confirming their responsibility for the 1.6~eV--PL band. 

\section{Discussion}

We observed different surface imperfections with similar PL features in WSe$_2$/MoS$_2$ and MoS$_2$/WSe$_2$ heterobilayers. In general, on SiO$_2$, only blisters are generated when the two layers are stacked, and they are distinguished by showing strong and redshifted PL of the top layer. We also detected rare protrusions caused by a nanoparticle on SiO$_2$ when the substrates were not clean enough (Figures S11-13) that generate local strains as in the case of gas or liquid-filled blisters. On an hBN substrate, nanobubbles and blisters are created in both layers regardless on the stacking order. When WSe$_2$ is on top, the PL of WSe$_2$ in these protrusions exhibits larger shifts. No signatures of IEX emission in the range of 1.3--1.6~eV was observed in MoS$_2$--WSe$_2$ heterobilayers.

Although the electronic nature of the IEX in the different systems is still under debate, \cite{Hanbicki2018,Wilson2017,Ponomarev2018,Okada2018,Kiemle2020} we focus our discussion mostly on the MoS$_2$-WSe$_2$ system. While the space indirect emission has been predicted \cite{Chiu2015,Zhang2017,Karni2019} and experimentally measured to be at energies close to 1~eV,\cite{Binder2019,Karni2019,Huang2020} the most common emission observed by $\mu$-PL at energies between 1.55--1.6~eV was identified as a momentum-space indirect owing to the large lattice mismatch and the offset of energy levels.\cite{Fang2014,Kunstmann2018,Nagler2019,Chiu2014,Unuchek2018,Ji2020,Lin2015} This particular interlayer emission was demonstrated to possess a weak twist angle dependence and became undetectable when the temperature is decreased.\cite{Kunstmann2018}

However, several factors made us consider a possible reassignment of the PL bands reported in this range.

1) The PL of the blisters and nanobubbles is much stronger than the PL of the flat areas, simply because the PL quenching is lifted due to the detachment of the layers by the encapsuled contamination, a possibility that has not been discussed in previous works. The strong PL intensity together with the strain-induced shift from these heterogeneities can influence the measurements. The lack of spatial resolution in the previous $\mu$-PL studies makes the distinction between locally corrugated and flat areas almost impossible. Moreover, Huang at el. \cite{Huang2020} recently demonstrated that the interlayer emission from MoS$_2$--WSe$_2$ lies at around 1~eV without any signal at around 1.6~eV. 

2) The momentum-space indirect interlayer emission was demonstrated to be independent of the stacking angle between layers.\cite{Kunstmann2018} Nevertheless, some researchers have assigned the interlayer emission at around 1.59~eV,\cite{Chiu2014,Kunstmann2018} while others observed it at around 1.55~eV\cite{Fang2014,Ji2020}. This matches well with the measured energy for blisters and nanobubbles, respectively, with the variation being the result of different strain levels.

3) It was also demonstrated that interlayer emission disappears at low temperatures\cite{Chiu2014,Kunstmann2018}, which is unsupported by other observations in MoS$_2$-WS$_2$ and MoSe$_2$-WSe$_2$,\cite{Hanbicki2018,Kiemle2020,Okada2018} where momentum indirect interlayer emission maintains its dependence on the temperature. However, we are aware that the lattice mismatch in MoS$_2$/WSe$_2$ may cause a different fundamental behavior.

4) As-grown CVD MoS$_2$-WSe$_2$ vertical heterobilayers are more difficult to synthesize than other vertical heterobilayers due to the high tendency to form alloys,\cite{Li2015} and not many authors have reported the PL signature of such systems. To the best of our knowledge, only one study reported the PL of an as-grown sample, assigning the 1.6~eV-PL band as the interlayer emission.\cite{Lin2015} However, the band at 1.6~eV is broad and accompanied by redshifts of both intralayer excitons (MoS$_2$ and WSe$_2$), making the fitting and subsequent band assignment problematic.\cite{Lin2015} More experiments in this line of study are sorely needed.

5) In general, the position of the monolayer (MoS$_2$ and WSe$_2$) in the stacking order should not affect the band energy alignment.\cite{Chiu2015} Yet, in our samples, the band at $\approx$~1.6~eV always appears at a higher energy in MoS$_2$/WSe$_2$. The absence of emission at around 1.6 eV when using MoS$_2$ as a top layer can be found in some works.\cite{Lee2014, Zhu2017} However, as discussed above, the blisters and nanobubbles that decouple the two layers provide the simultaneously enhanced and shifted PL for the top layer. Therefore, the TEPL signal of WSe$_2$ shifts---or broadens in $\mu$-PL---toward lower energies only when this particular layer is on top.

\section{Conclusions}

To conclude, we have investigated different combinations of heterobilayers composed of MoS$_2$, MoSe$_2$, WS$_2$, and WSe$_2$ on SiO$_2$ and hBN, with tip-enhanced PL and high resolution AFM. We show that while MoS$_2$-WS$_2$ and MoSe$_2$-WSe$_2$ exhibit a clear signature of interlayer excitons in the energy range measured previously by PL with diffraction-limited resolution, the TEPL spectra of the MoS$_2$-WSe$_2$ heterostructure do not contain any clear sign of the IEX emission in the often reported energy range. Instead, we evidence PL shifts and intensity changes caused by local topography features, such as nanobubbles and blisters, which are formed between the layers due to contamination. The IEX-suspicious PL shift is absent even when the two layers exhibit a Moiré pattern, which proves a strong interaction between them. In contrast, such PL shifts take place in an optically flat sample even when a single nanobubble detectable by AFM is present. Additional effects of various nanobubble configurations on the PL spectra are described, highlighting the need for a careful interpretation of the $\mu$-PL spectra of van der Waals heterostructures.  

\section{Methods}

\subsection{Sample preparation}
The samples were prepared by mechanical exfoliation of bulk crystals (2D semiconductors, HQ graphene). The monolayers were separately exfoliated on polydimethylsiloxane (PDMS) stamps and transferred onto SiO$_2$/Si substrates (300 nm SiO$_2$) using a dry-transfer technique.\cite{Castellanos_2014} For the samples on hBN, we first transferred 15--20 nm-thick hBN flakes onto the substrates using the same method. The temperature was set to 70\textdegree~C after contacting the PMDS stamp on the substrate and cooled down before releasing the contact. The temperature facilitates the interaction between the layers and interface self-cleaning, which is evidenced by the creation of blisters. \cite{Kretinin2014}

\subsection{Micro- and nano-photoluminescence characterization}
The micro PL measurements were performed in a LabRAM HR Evolution spectrometer (Horiba Scientific) with a 633~nm laser excitation and 150 l/mm diffraction grating. The TEPL measurements were performed with the same LabRAM HR Evolution spectrometer coupled to an OmegaScope SPM (Horiba Scientific) using side-illumination with 633~nm laser excitation, 150 l/mm grating, and a 100x objective (0.7 NA). Own-made plasmonic tips were used by sputtering Ag on Si probes (Access-fm, App Nano). TEPL imaging was performed by using DualScan mode, where the AFM feedback switches from contact to semi-contact at each pixel of the scan. The emission collected when the tip is in contact contains both the near-field and far-field signals, while the far-field signal dominates the emission when the tip is in the semi-contact mode. The TEPL signal was obtained after the pixel by pixel subtraction of the far-field from the map that was collected in the contact mode (Figure S6). The applied contact force was weak (5~nN),\cite{Darlington_1_2020} as we used the minimum force needed to maintain stable feedback during the scan. The acquisition time for each pixel was 0.2--0.5~s for TEPL and 1s for $\mu$-PL maps, unless specified otherwise. The $\mu$-PL single spectra were collected with longer acquisition times. For all the measurements, the laser power was set between 40 and 200~$\mu$W.

\subsection{Topography characterization}
The AFM images were taken with a Bruker Dimension ICON in PeakForce Tapping mode using Scanasyst Air probes (Bruker Corp).

\begin{acknowledgement}

This work was funded by the Czech Science Foundation (GACR 20-08633X). The study was further supported by the Pro-NanoEnviCz project (Reg. No. CZ.02.1.01/0.0/0.0/16\_013/\allowbreak0001821) supported by the Ministry of Education, Youth and Sports of the Czech Republic and the European Union - European Structural and Investments Funds in the frame of Operational Programme Research Development and Education. 

\end{acknowledgement}

\begin{suppinfo}

Additional $\mu$-PL maps of MoS$_2$-WSe$_2$ heterobilayers, $\mu$-PL spectra of different heterobilayers showing the IEX emission, TEPL details of MoS$_2$-WSe$_2$ and MoSe$_2$/WSe$_2$ heterobilayers, and TEPL maps of protrusions originated by nanoparticles in MoS$_2$-WSe$_2$ heterobilayers.

\end{suppinfo}

\bibliography{HeteroTEPL}

\newpage
\section*{Graphical TOC Entry}
\includegraphics{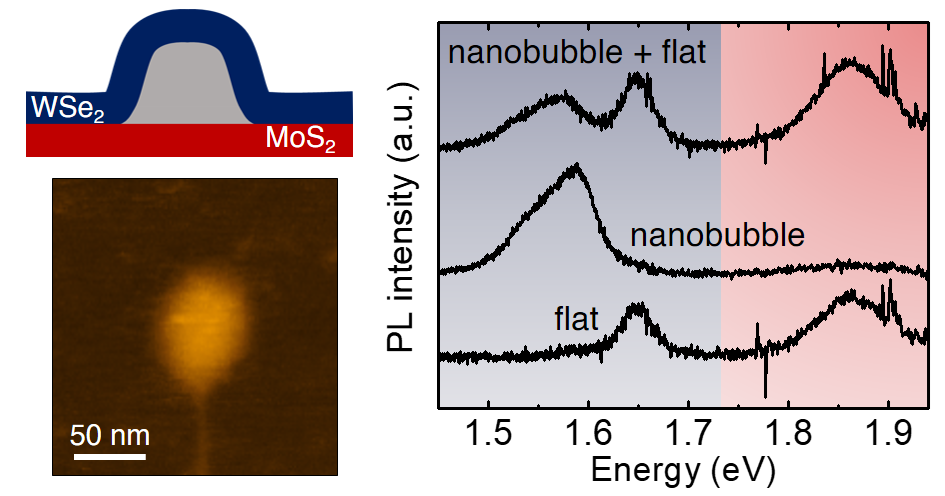}

\end{document}

\newpage

\section{Micro-photoluminescence of heterobilayers on SiO$_2$}
    \subsection{WSe$_2$/MoS$_2$ on SiO$_2$.}
    
\begin{figure}
    \includegraphics[width=0.6\textwidth]
    {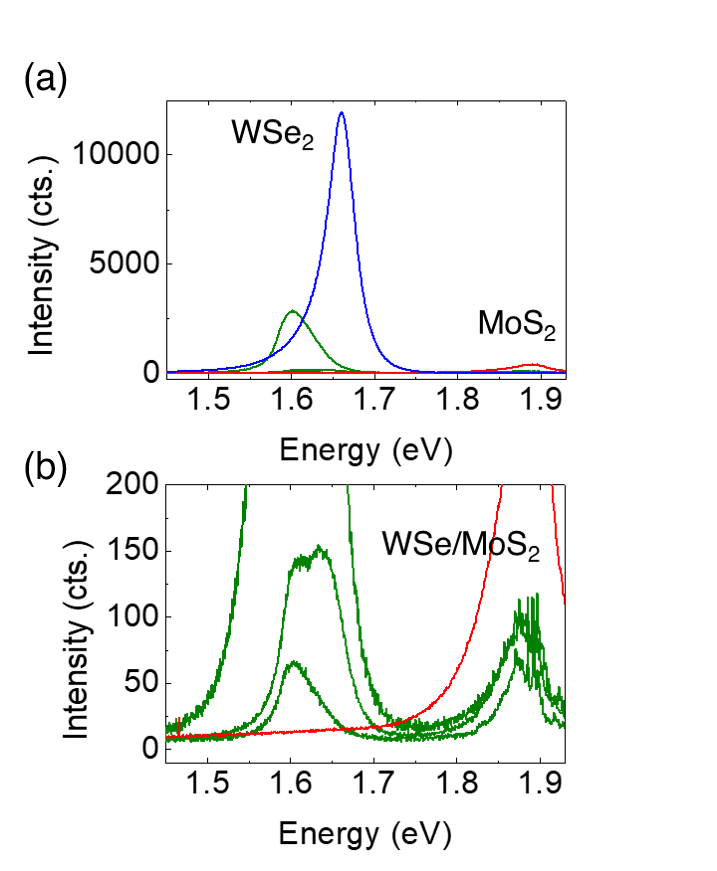}
    \caption{(a) Absolute intensity $\mu$-PL spectra of the WSe$_2$/MoS$_2$ heterobilayer shown in Figure 1. (b) Enlarged region focusing on the heterobilayer spectra.}
    \label{fgr:S1}
\end{figure}

\begin{figure}
 \includegraphics[width=0.8\textwidth]
 {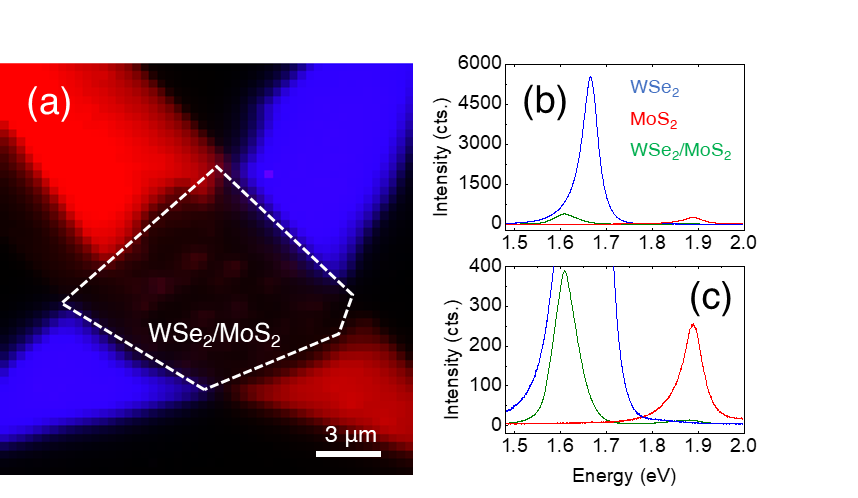}
  \caption{(a) The $\mu$-PL intensity map of WSe$_2$/MoS$_2$ (sample 2), laser excitation 532 nm. (b) Absolute intensity $\mu$-PL spectra of the same WSe$_2$/MoS$_2$ heterobilayer. (c) Enlarged region.}
  \label{fgr:S2}
\end{figure}

Figure \ref{fgr:S1} shows the absolute intensity PL spectra of MoS$_2$ and WSe$_2$ monolayers, and a heterobilayer when the WSe$_2$ monolayer is placed on top. The spectra are the same as the ones shown in Figure 1(c), which are normalized to allow their comparison. The PL is extremely quenched in the heterobilayer area. The enlarged area shown in Figure \ref{fgr:S1}(b) illustrates the variation of the PL across the sample. A PL map and representative spectra of a different sample are shown in Figure \ref{fgr:S2}, confirming the reproducibility of the results.

    \hspace{\fill} \\

   \subsection{MoS$_2$/WSe$_2$ on SiO$_2$}
    
\begin{figure}
 \includegraphics[width=\textwidth]
 {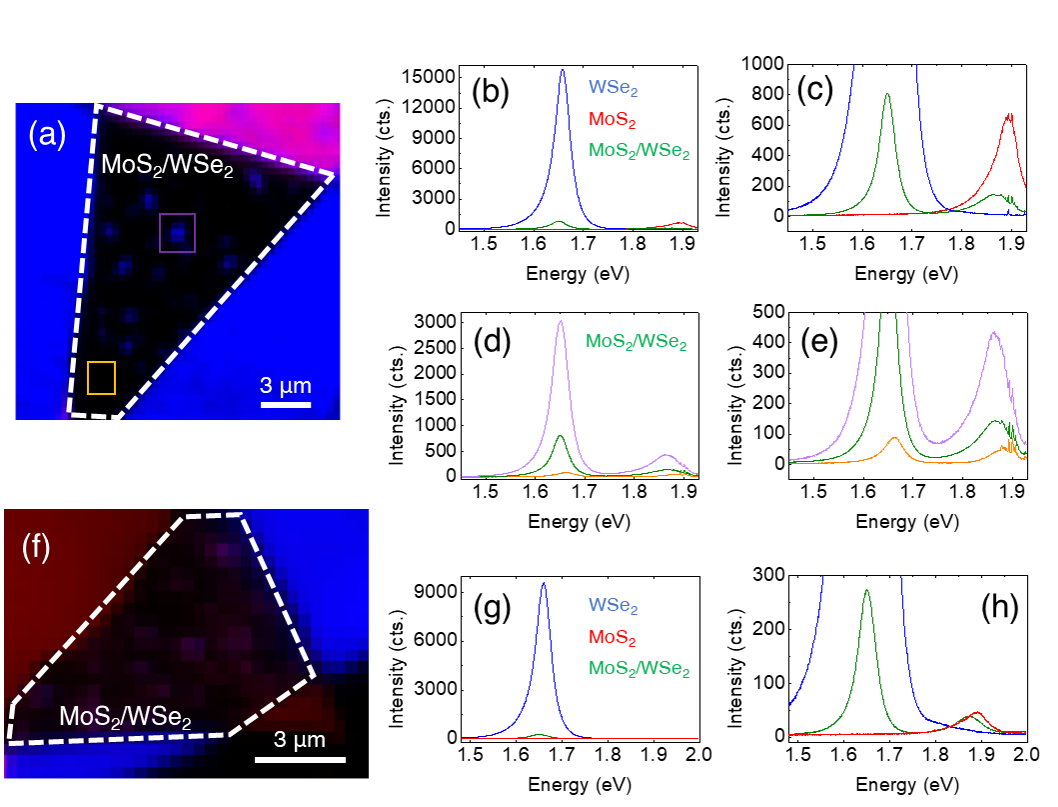}
  \caption{$\mu$-PL characterization of MoS$_2$/WSe$_2$ heterobilayers. (a) PL intensity map of MoS$_2$/WSe$_2$ (sample 1), laser excitation 633 nm. (b) Absolute intensity $\mu$-PL spectra of MoS$_2$/WSe$_2$. (c) Enlarged region. (d) Absolute intensity $\mu$-PL spectra of MoS$_2$/WSe$_2$ taken in different areas, as indicated in (a) with matching colors. The green spectrum is the average of the whole area. (f) PL intensity map of MoS$_2$/WSe$_2$ (sample 2), laser excitation 532 nm (g) Absolute intensity $\mu$-PL spectra of MoS$_2$/WSe$_2$. (h) Enlarged region.}
  \label{fgr:S3}
\end{figure}

Figure \ref{fgr:S3} (a-c) shows the PL intensity map and the absolute intensity PL spectra of MoS$_2$ and WSe$_2$ monolayers, and a heterobilayer when the MoS$_2$ is placed on top. The PL is extremely quenched in the heterobilayer area. Selected PL spectra are shown in Figure \ref{fgr:S3} (d-e) where the green spectrum corresponds to the average of the whole heterobilayer, while orange and purple ones to the average areas indicated in Figure \ref{fgr:S3} (a).
For this stacking order, the spectra do not show any band at around 1.6 eV, not even in the most quenched area (orange spectrum) where the strongest interlayer coupling is expected. The PL spectra of a second sample is shown in Figure \ref{fgr:S3} (f-h), confirming the reproducibility of the results.

    \clearpage
    \subsection{Other heterobilayers}
    
\begin{figure}
     \includegraphics[width=\textwidth]
     {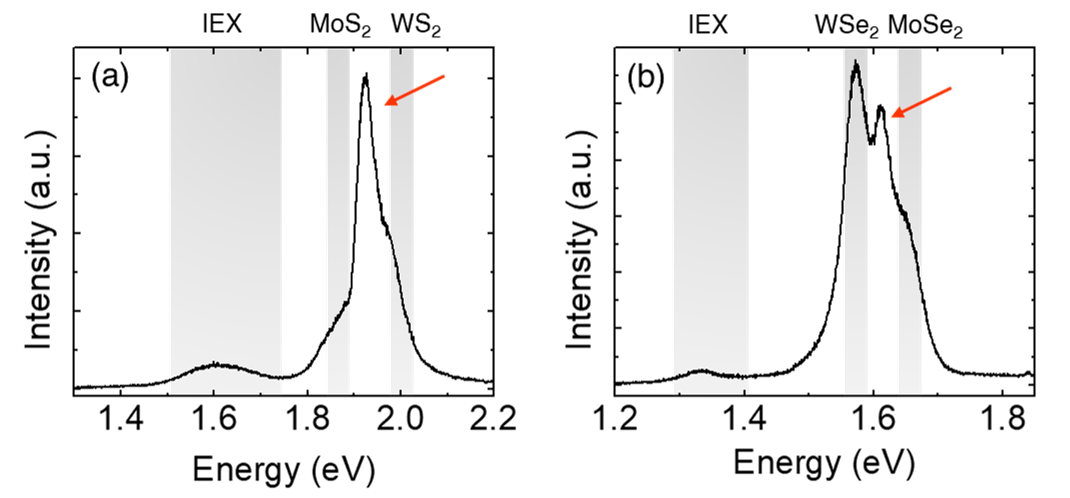}
     \caption{The $\mu$-PL spectra of WX$_2$/MX$_2$ heterobilayers with WX$_2$ layer on top. (a) WS$_2$/MoS$_2$, (b)WSe$_2$/MoSe$_2$.}
      \label{fgr:S8}
    \end{figure}

\begin{figure}
     \includegraphics[width=\textwidth]
     {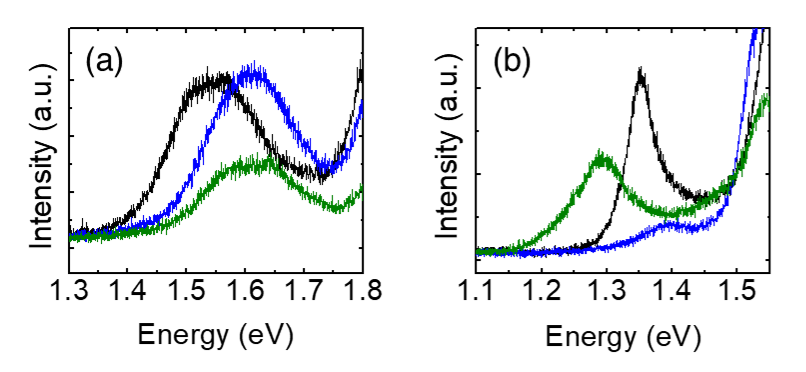}
     \caption{The $\mu$-PL spectra of different (a) MoS$_2$/WS$_2$, (b) MoSe$_2$/WSe$_2$ heterobilayers showing the interlayer emission at different energies.}
    \label{fgr:S9}
    \end{figure}

    Figure \ref{fgr:S8} shows the $\mu$-PL spectra of the heterobilayers when WS$_2$ and WSe$_2$ are placed on top. The PL spectra show a third peak (indicated with a red arrow) between the intralayer emission of the monolayers that may be assigned to the blisters  by analogy to the previously measured systems. Although the samples contain blisters, the interlayer emission can be measured with $\mu$-PL since the areas between blisters exhibit a good interaction. Similar features were reported before in WS$_2$/MoS$_2$ heterobilayers and the extra peak was defined as a P$_{het}$.\cite{Tongayb2014} 
    We observed the interlayer emission in different samples with both the W and Mo-containing layer on top in both systems. Figure \ref{fgr:S9} shows the PL interlayer emission of different samples. The PL position and intensity depends on the stacking angle and cleanliness of the interface.
    \newline
    
 \hspace{\fill}\\
 \vspace*{1.5cm}
 
    \section{Tip-enhanced photoluminescence of heterobilayers on SiO$_2$ and hBN}

    \subsection{WSe$_2$/MoS$_2$ on SiO$_2$}
    
\begin{figure}
 \includegraphics[width=\textwidth]
 {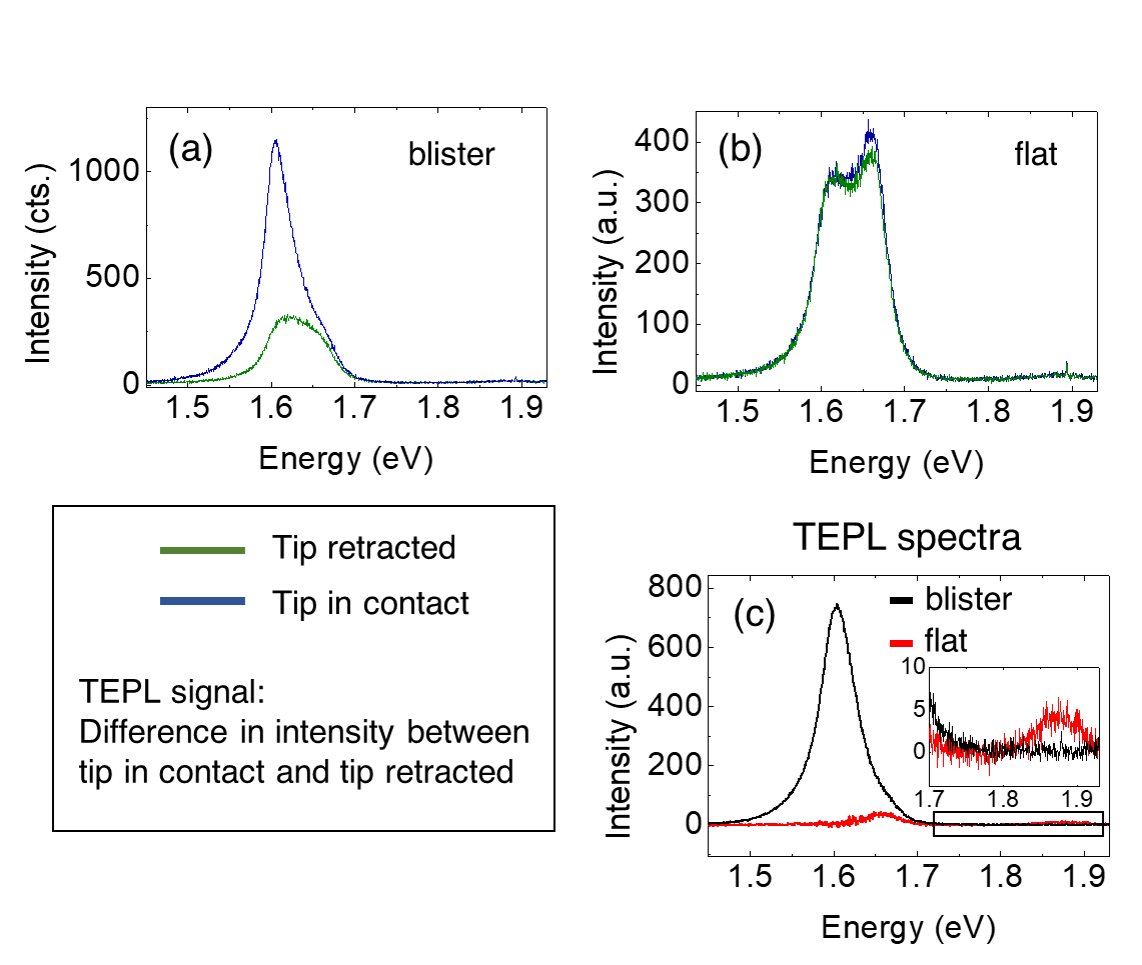}
  \caption{TEPL measurement of WSe$_2$/MoS$_2$ heterobilayer. (a) PL spectra when the tip is in contact (blue line) and retracted (green line) from a blister. (b) PL spectra when the tip is contacted and retracted in a flat area. (c) Absolute intensity TEPL spectra obtained by the subtraction of the previous spectra in a blister (black spectrum) and in a flat area (red spectrum). The inset shows the enlarged region indicated in the figure.}
  \label{fgr:S4}
  \end{figure}

Figure \ref{fgr:S4} compares the spectra taken when the tip is in contact and when the tip is retracted (semicontact mode). The TEPL signal is obtained by subtracting the retracted spectrum (far-field) from the spectrum taken in contact. An example of the stronger enhancement observed in the blisters is also illustrated in Figure \ref{fgr:S4}. The subtracted TEPL spectra are shown in the Figure \ref{fgr:S4}(c); they correspond to the normalized spectra shown in Figure 1 (main text).

\clearpage
    
    \subsection{MoS$_2$/WSe$_2$ on SiO$_2$}
    
\begin{figure}
 \includegraphics[width=0.9\textwidth]
 {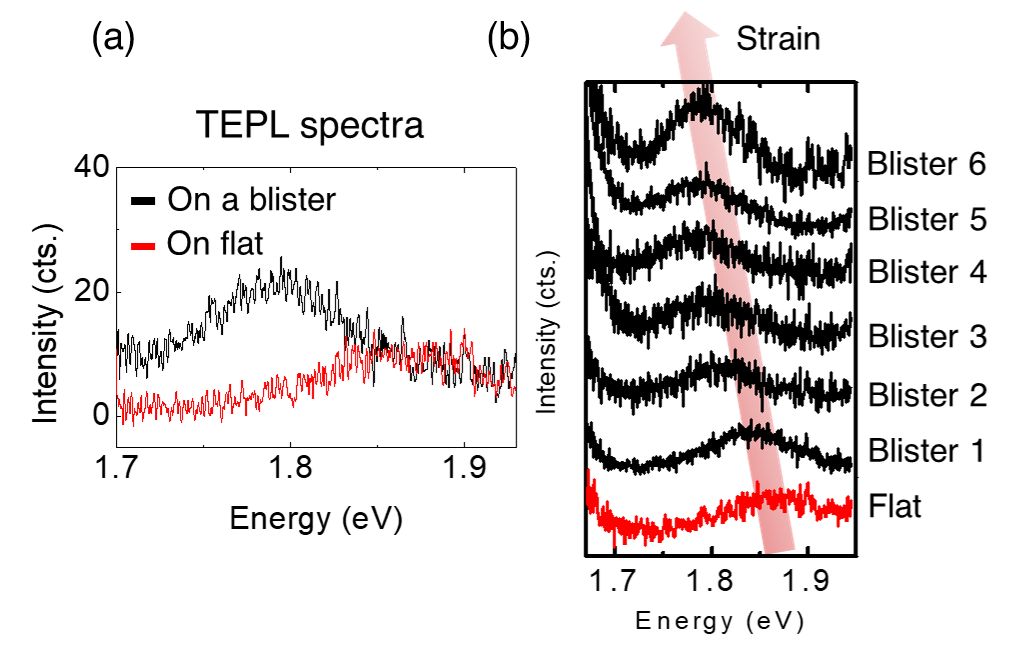}
  \caption{\textbf{TEPL spectra of MoS$_2$ blisters in MoS$_2$/WSe$_2$ heterobilayers.} (a) Absolute intensity TEPL of a blister and of a flat area. (b) TEPL spectra of different MoS$_2$ blisters.}
  \label{fgr:S5}
\end{figure}

Figure \ref{fgr:S5}(a) shows the non-normalized TEPL spectra from Figure 1(k) (main text) in the MoS$_2$ exciton energy range. The enhancement is one order of magnitude weaker than the one observed in WSe$_2$ blisters (Figure \ref{fgr:S4}(c). Different blisters exhibits different PL shift as depicted in Figure \ref{fgr:S5}(b). 

     \subsection{MoS$_2$/WSe$_2$ on hBN}
     
\begin{figure}
 \includegraphics[width=0.7\textwidth]
 {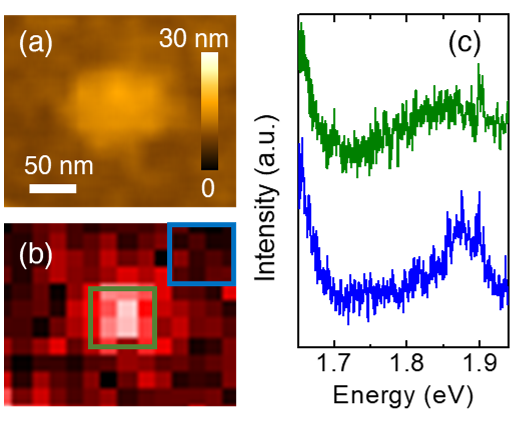}
  \caption{TEPL spectra of an MoS$_2$ nanobubble in MoS$_2$/WSe$_2$ on hBN. (a) AFM image. (b) TEPL map. (c) Average TEPL spectra of the areas indicated in (b).}
  \label{fgr:S13}
\end{figure}

We also observed MoS$_2$ nanobubbles in heterobilayers transferred on hBN when MoS$_2$ was placed on top of WSe$_2$. Figure \ref{fgr:S13} shows the topography and TEPL intensity map of a MoS$_2$ nanobubble. Although the enhancement is weaker than the one observed in WSe$_2$ nanobubbles, it can be seen in the spectra (Figure \ref{fgr:S13}(c)) that the PL of MoS$_2$ intralayer exciton shifts to lower energies in the nanobubble area. 
     
\vspace*{0.5cm}
    \subsection{MoSe$_2$/WSe$_2$ on SiO$_2$}
    
\begin{figure}
 \includegraphics[width=0.4\textwidth]
 {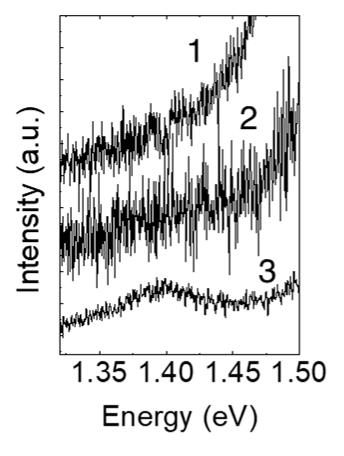}
  \caption{Enlarged TEPL spectra of MoSe$_2$/WSe$_2$ heterobilayers over the interlayer exciton emission energy (\textit{cf}. Figure 4(f), main text).}
  \label{fgr:S6}
\end{figure}

\begin{figure}
 \includegraphics[width=0.8\textwidth]
 {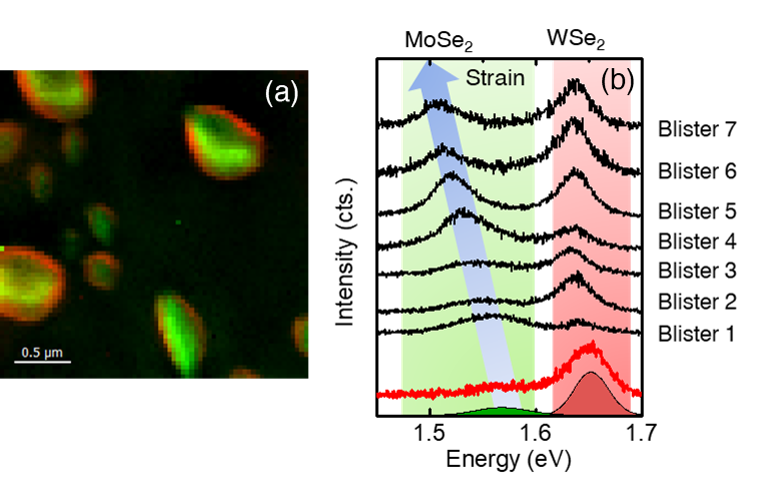}
  \caption{TEPL spectra of MoSe$_2$/WSe$_2$ heterobilayers. (a) TEPL map of a MoSe$_2$/WSe$_2$ heterobilayer. The blisters are formed by strained MoSe$_2$. (b) TEPL spectra of different MoSe$_2$ blisters.}
  \label{fgr:S7}
\end{figure}

An enlarged area of the Figure 4(f) of the main text is shown in Figure \ref{fgr:S6}. The difference in the signal/noise ratio is due to the different number of pixels used in the averaging. In Figure \ref{fgr:S7}, a TEPL intensity map and selected spectra of MoSe$_2$/WSe$_2$ are displayed. The color map corresponds to the integrated intensity of the regions indicated in the graph with matching colors. As in the previous systems, the blisters exhibit a shift of the PL constituting the top layer (MoSe$_2$).

\section{Protrusions formed by a nanoparticle.}
\begin{figure}
 \includegraphics[width=0.8\textwidth]
 {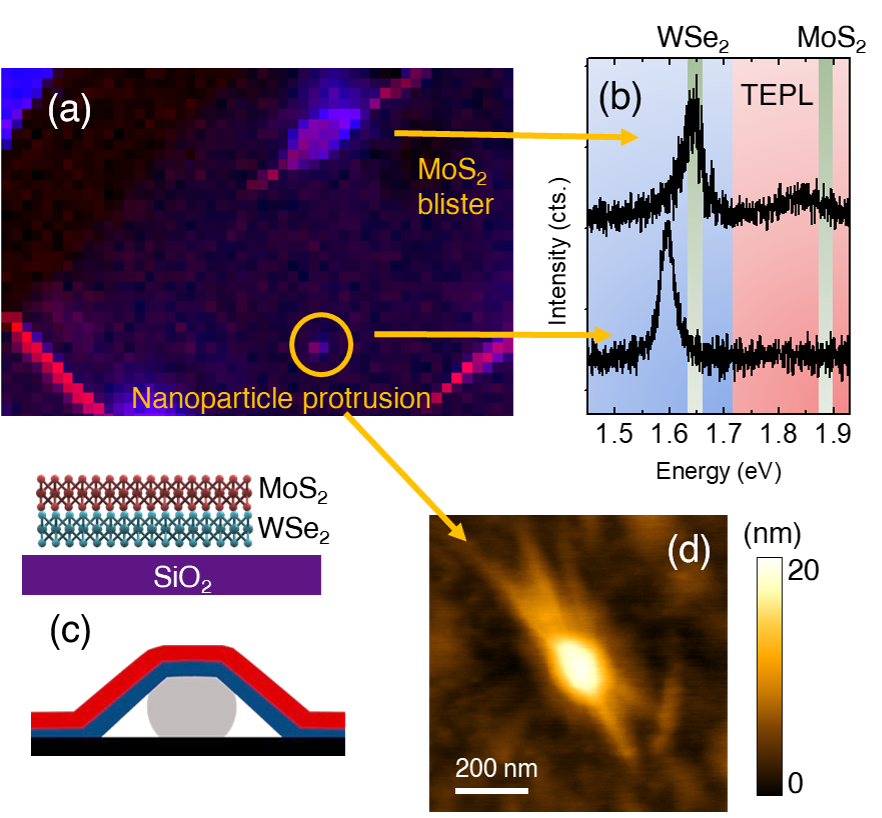}
  \caption{Protrusion generated by a nanoparticle in MoS$_2$/WSe$_2$ heterobilayer showing PL around the reported interlayer exciton emission energy. (a) TEPL spectra with a typical blister showing PL of the top layer (MoS$_2$) shifted and an additional area with a different PL signature. (b) PL spectra of the areas indicated in (a). (c) Topography image of the area indicated in (a).}
  \label{fgr:S10}
\end{figure}

Besides the PL emission from the blisters, we also found areas where a PL band near 1.6~eV appeared in MoS$_2$/WSe$_2$ heterobilayers on SiO$_2$ substrates. These areas were formed by protrusions generated by a nanoparticle below the layers. In the WSe$_2$/MoS$_2$ stack, these locations are difficult to detect since they present very similar PL signatures to the WSe$_2$ blisters. In MoS$_2$/WSe$_2$---since the samples show no shift of the WSe$_2$ PL---we were able to find them easily (Figure \ref{fgr:S10}). In the TEPL map, one can see areas where the PL of MoS$_2$ is shifted due to the formation of blisters (Figures 1 and \ref{fgr:S5}); however, an area with PL emission at $\approx$~1.6~eV is also detected. By TEPL and AFM we can see that it is actually a protrusion formed by a nanoparticle---the topography shows clear wrinkles emanating from the lifted area.

\begin{figure}
 \includegraphics[width=\textwidth]
 {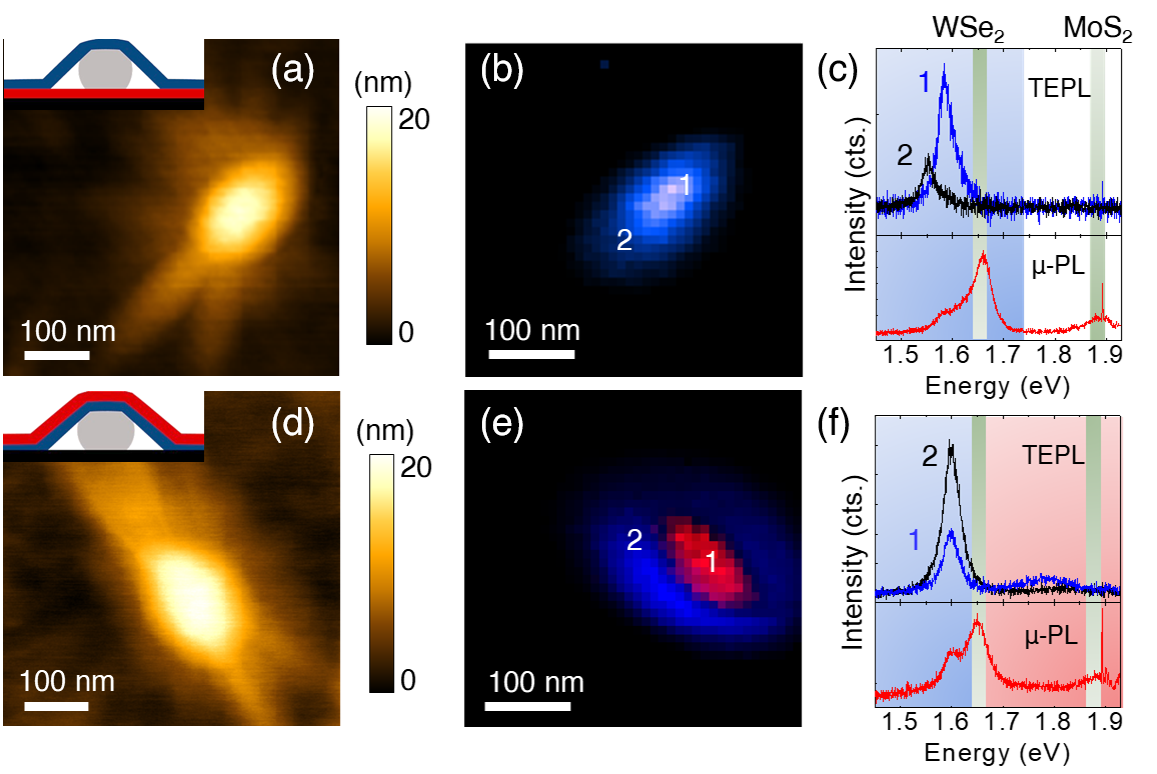}
  \caption{Comparison between WSe$_2$/MoS$_2$ and MoS$_2$/WSe$_2$ when a nanoparticle is below the layers. Topography images of (a) WSe$_2$/MoS$_2$ and (d) MoS$_2$/WSe$_2$ heterobilayers. TEPL intensity maps of (b) WSe$_2$/MoS$_2$ and (e) MoS$_2$/WSe$_2$ heterobilayers. The colors refer to the integrated intensity of the areas indicated in (c) and (f). Selected TEPL spectra of (c) WSe$_2$/MoS$_2$ and (f) MoS$_2$/WSe$_2$ heterobilayers of the areas marked in (b) and (e); $\mu$-PL spectra of the respective areas are shown for comparison.}
  \label{fgr:S11}
\end{figure}

\begin{figure}
 \includegraphics[width=\textwidth]
 {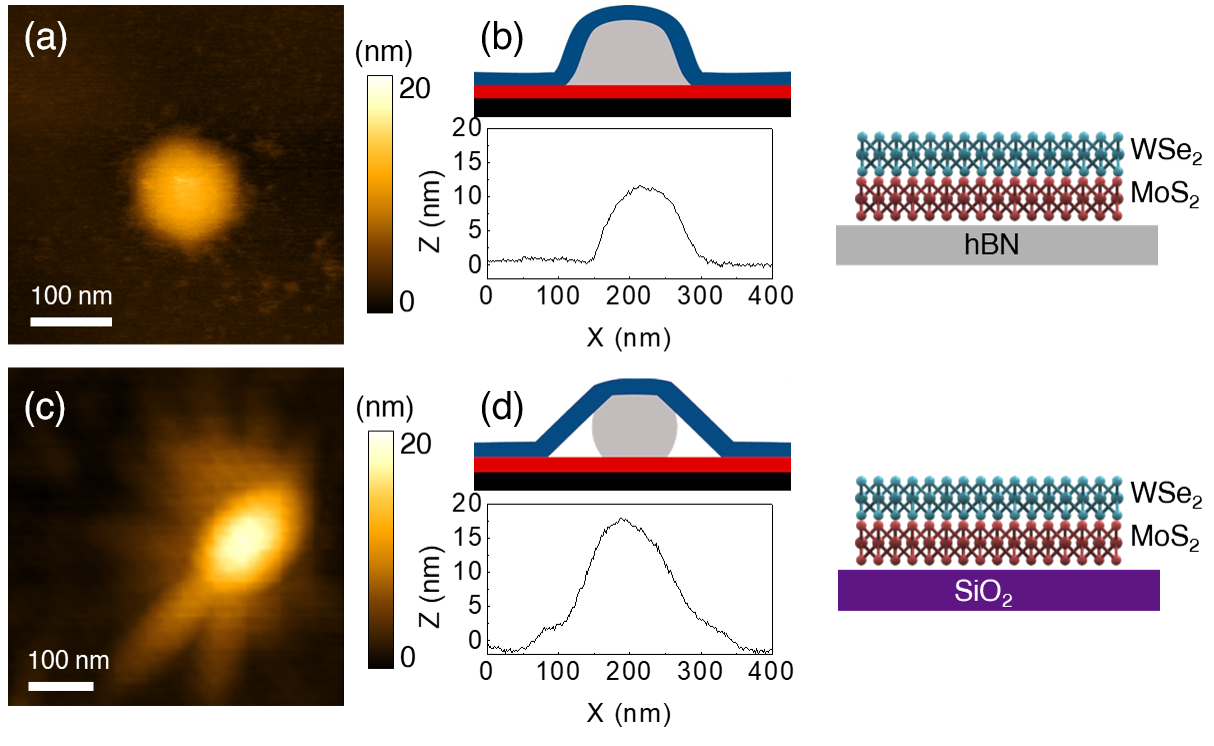}
  \caption{Comparison between nanobubbles formed by gas/liquid and protrusions generated by a nanoparticle below. We have observed gas/liquid nanobubbles in the samples when using hBN as a substrate. The protrusions with a particle below are more rare and can be observed on SiO$_2$ substrates. Topography images of (a) nanobubble observed in WSe$_2$/MoS$_2$ on hBN, (b) protrusion on a nanoparticle in WSe$_2$/MoS$_2$ on SiO$_2$.}
  \label{fgr:S12}
\end{figure}

A comparison of protrusions formed in WSe$_2$/MoS$_2$ and MoS$_2$/WSe$_2$ is shown in Figure \ref{fgr:S11}. In the TEPL analysis, one can see that the distribution of the PL is different for the two configurations and similar to the distribution observed in nanobubbles formed on hBN (Figure 3, main text). On the one hand, when WSe$_2$ is on top (Figure \ref{fgr:S11}(a-c)), the PL of WSe$_2$ is redshifted, showing the same energy distribution as the nanobubbles, where the PL at the periphery is more shifted but less intense. This signature provoked by strain may be mixed up with the reported IEX,\cite{Fang2014,Kunstmann2018,Ji2020} as observed in the $\mu$-PL spectrum of a nearby area. On the other hand, when MoS$_2$ is placed on top (Figure \ref{fgr:S11}(d-f)), the PL of both WSe$_2$ and MoS$_2$ is redshifted in the central areas, while only the PL of the WSe$_2$ is redshifted at the periphery. This signature may also be interpreted as the IEX emission, when looking at the $\mu$-PL spectrum. These protrusions exhibit very similar strain distributions as the nanobubbles observed in the samples fabricated on hBN.\cite{Darlington_1_2020} However, the AFM topography evidences the different physical nature between them. Figure \ref{fgr:S12} compares both types of features when WSe$_2$ layer is on top. While the height profile of a nanobubble forms an almost perfect symmetric sinusoidal shape, the protrusion induces an uneven partial suspension of the layer, resulting in an asymmetric topography with more inflection points.

\bibliography{HeteroTEPL_Sup}